\documentclass{iau_FM}
\usepackage{graphicx}
\usepackage{subfig}
\usepackage{multicol}

\title[Molecular gas in distant LLRGs] 
{Molecular gas in radio galaxies at $z=0.4-2.6$ in (proto-)cluster environment} 

\author[G. Castignani et al.]   
{G.~Castignani$^{1,2,3}$,
F.~Combes$^{2,3}$,
P.~Salom\'{e}$^2$,
C.~Benoist$^4$,
M.~Chiaberge$^{5,6}$,
J.~Freundlich$^7$,
\and G.~De Zotti$^8$}

\affiliation{$^1$ Laboratoire d'astrophysique, \'{E}cole Polytechnique F\'{e}d\'{e}rale de Lausanne (EPFL), Observatoire de Sauverny, 1290 Versoix, Switzerland \\ email: {\tt gianluca.castignani@epfl.ch} \\[\affilskip]
$^2$Sorbonne Universit\'{e}, Observatoire de Paris, PSL, CNRS, LERMA, F-75014, Paris, France \\
$^3$Coll\`{e}ge de France, 11 Place Marcelin Berthelot, 75231 Paris, France\\
$^4$Universit\'{e} C\^{o}te d'Azur, Observatoire de la C\^{o}te d'Azur,  CNRS, Laboratoire Lagrange, Blvd de l'Observatoire, CS 34229, 06304 Nice cedex 4, France \\
$^5$Space Telescope Science Institute, 3700 San Martin Dr., Baltimore, MD 21210, USA\\
$^6$Johns Hopkins University, 3400 N. Charles Street, Baltimore, MD 21218, USA \\
$^7$Centre for Astrophysics and Planetary Science, Racah Institute of Physics, The Hebrew University, Jerusalem 91904, Israel\\
$^8$INAF-Osservatorio Astronomico di Padova, Vicolo dell'Osservatorio 5, I-35122 Padova, Italy}
	     
\pubyear{2018}
\jname{Astronomy in Focus, Volume 1} 
\editors{Volker Beckmann \& Claudio Ricci, eds.}
\begin{document}
\maketitle

\vspace{-0.2cm}
\begin{abstract}
We investigate the role of the environment in processing molecular gas in radio galaxies (RGs). We observed five RGs at $z=0.4-2.6$ in dense Mpc-scale environment with the IRAM-30m telescope. We set four upper-limits and report a tentative CO(7$\rightarrow$6) detection for COSMOS-FRI~70 at $z=2.63$, which is the most distant brightest cluster galaxy (BCG) candidate detected in CO. 
We speculate that the cluster environment might have played a role in preventing the refueling via environmental mechanisms such as galaxy harassment, strangulation, ram-pressure, or tidal stripping.
The RGs of this work are excellent targets for ALMA as well as next generation telescopes such as the {\it James Webb Space Telescope. }
\keywords{Galaxies: active; Galaxies: clusters: general; Molecular data.}
\end{abstract}

\noindent{\bf 1. Introduction.} Molecules in galaxies can help to trace star forming regions, even close to an active galactic nucleus (AGN, Omont 2007). 
Radio galaxies (RGs) are  a precious tool to discover distant galaxy groups and (proto-)clusters at high-$z$, since they are often the brightest cluster galaxies  (BCGs, Zirbel 1996). 

We consider two RGs at $z = 0.39$ and 0.61 within the DES SN deep fields (DES collaboration 2015) and three additional COSMOS-FRI RGs from the Chiaberge et al. (2009) sample, at $z=0.97$, 0.91, and 2.63. They have been selected since they show evidence of significant star formation, SFR$_{24\mu m}\sim(20-250)M_\odot$/yr, based on their emission at $\sim24\mu$m in the observer frame (WISE or Spitzer-MIPS). 
All five sources are found  in (proto-)cluster candidates by using the Poisson Probability Method (PPM, Castignani et al. 2014ab) that searches for overdensities using photometric redshifts.
\smallskip

\begin{figure}[h!]
\begin{center}
\vspace{-1.cm}
\subfloat{\includegraphics[width=0.33\textwidth]{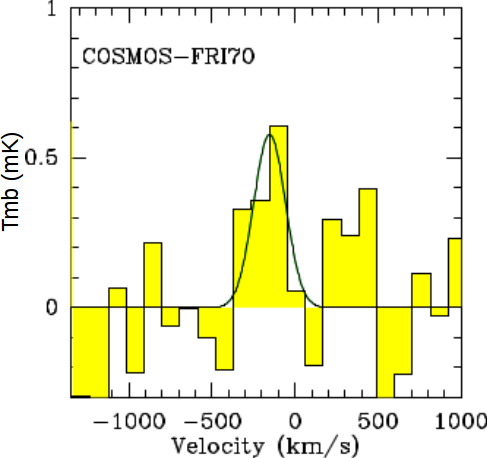}}
\subfloat{\hspace{0.1cm}\raisebox{0.7cm}{\includegraphics[width=0.23\textwidth]{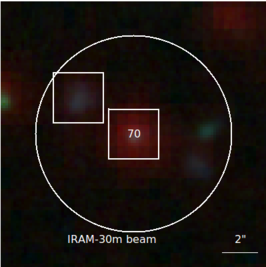}}}
\subfloat{\hspace{0.1cm}\includegraphics[width=0.4\textwidth]{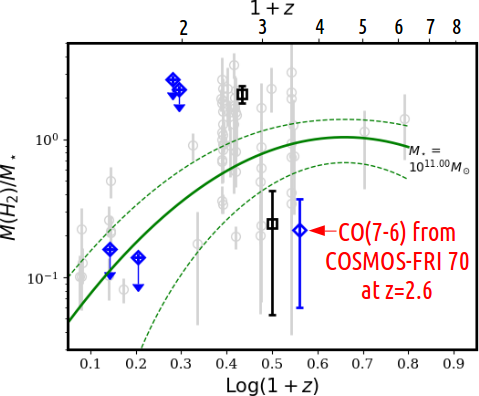}}\\
\end{center}
\caption{Left: CO(7-6) spectrum of COSMOS-FRI~70 taken with IRAM-30m. Center: RGB image of COSMOS-FRI~70 and its nearby companions, North is up, East is left. Right: molecular gas properties of BCG candidates from our IRAM-30m campaign are highlighted as blue diamonds, BCGs from other studies are shown as black squares (Emonts et al. 2013, Webb et al. 2017), gray circles denote normal cluster galaxies, model predictions from Tacconi et al. (2018) for MS galaxies and their uncertainties are shown as green solid and dashed lines, respectively.}\label{fig:figure}
\end{figure}

\noindent{\bf 2. Methods and Results.}
We observed the five targets with the IRAM-30m telescope targeting several CO(J$\rightarrow$J-1) lines, one for each source, at $\sim$(1.2-1.4)~mm in the observer frame (Castignani et al. 2018). Data reduction and analysis were performed using the CLASS software of the GILDAS package.\footnote{https://www.iram.fr/IRAMFR/GILDAS/} Our observations yielded four CO upper limits and one tentative CO(7-6) detection for COSMOS-FRI~70, which makes it the most distant BCG candidate detected in CO. There are in fact pieces of evidence that the radio source is { the BCG of one of the most distant proto-clusters hosting a RG.} i)~The stellar mass is exceptionally high ($\log(M_\star/M_\odot)\simeq11$, Baldi et al. 2013). ii)~By applying Galfit (Peng et al. 2002) to the archival high-resolution HST ACS F814W (I-band) image a S\`{e}rsic index $4.3\pm0.6$ is found, consistently with that of early type galaxies. iii)~Color-color and color-magnitude plots further suggest that COSMOS-FRI~70 is indeed a { star forming massive elliptical}, $\sim0.3$mag brighter than all photometrically selected (proto-)cluster members (Castignani et al. 2018). Our results are reported in  Table~\ref{tab:radio_galaxies_properties_mol_gas} and Fig.~\ref{fig:figure}.

\smallskip
\noindent{\bf 3. Conclusions.} All target RGs have molecular gas properties that are consistent with the predictions for main sequence (MS) field galaxies. However they also show that high-$z$ BCGs i) tend to be gas poor and ii) have a relatively short depletion time scale. We thus speculate that the cluster environment might have played a role in preventing the refueling via environmental mechanisms such as galaxy harassment, strangulation, ram-pressure, or tidal stripping.

\begin{table}[h!]\centering
\vspace*{-0.3cm}
\begin{tabular}{ccccccccccccc}
\hline
 Galaxy ID &  $z_{spec}$ & CO(J$\rightarrow$J-1)  & $S_{\rm CO(J\rightarrow J-1)}$   &  $M({\rm H_2})$ & $\tau_{\rm dep}$ & $\frac{M({\rm H_2})}{M_\star}$  & $\tau_{\rm dep, MS}$   \\
   &  &  &  (Jy~km~s$^{-1}$)  & ($10^{10}~M_\odot$) & ($10^9$~yr) & ($10^8~M_\odot$) & ($10^9$~yr)  \\ 
 (1) & (2) & (3) & (4) & (5) & (6) & (7) & (8) \\
 \hline
{\small DES-RG~399} &    {\small 0.388439} & 3$\rightarrow$2   & $<1.5$ & $<$1.0 & $<$0.91 & $<$0.16  &  $1.10^{+0.16}_{-0.14}$ \\
 {\small DES-RG~708} &   {\small 0.60573} & 3$\rightarrow$2  & $<1.5$ & $<$2.6 & $<$0.36 & $<$0.14   &  $1.12^{+0.21}_{-0.17}$ \\
 {\small COSMOS-FRI~16} &  {\small 0.9687} & 4$\rightarrow$3 & $<5.5$ & $<$18.8 & $<$0.58 & $<$2.29 &  $0.91^{+0.16}_{-0.14}$  \\
 {\small COSMOS-FRI~31} &  {\small 0.9123} & 4$\rightarrow$3  & $<5.2$ & $<$15.8 & --- & $<$2.71   & $0.90^{+0.15}_{-0.13}$  \\
 {\small COSMOS-FRI~70} &  {\small 2.625} & 7$\rightarrow$6  & $0.69\pm0.31$ & $5.0\pm2.2$ & $0.20^{+0.13}_{-0.19}$ & $0.22^{+0.15}_{-0.16}$ & $0.66^{+0.17}_{-0.13}$ \\
 \hline
\end{tabular}
\vspace{-0.3cm}\caption{(1) galaxy name;  (2) spectroscopic redshift; (3-4) CO(J$\rightarrow$J-1) transition and flux; (5) molecular gas mass; (6) depletion time scale  $\tau_{\rm dep}=M({\rm H_2})/{\rm SFR_{24\mu m}}$; (7) molecular gas to stellar mass ratio; (8) $\tau_{\rm dep}$ predicted for MS field galaxies by Tacconi et al. (2018).}
\label{tab:radio_galaxies_properties_mol_gas}
\end{table}

\vspace{-0.6cm}
\begin{multicols}{2}

\end{multicols}

\end{document}